\documentclass[aps,prx,showpacs,preprintnumbers,twocolumn,superscriptaddress,longbibliography]{revtex4-1}
\usepackage{graphicx}
\usepackage{amssymb}
\usepackage{natbib}
\newcommand{\tr}{\mathrm{Tr}}
\usepackage{amsmath}
\usepackage{bbold}
\usepackage[caption=false]{subfig}
\usepackage{algorithm}
\usepackage{algpseudocode}
\usepackage{appendix}
\usepackage[dvipsnames]{xcolor}
\usepackage[section]{placeins}
\definecolor{darkblue}{rgb}{0,0,.65}
\definecolor{darkgreen}{rgb}{0.28,0.41,0.19}
\usepackage{wrapfig}
\usepackage[%
pdfstartview=FitH,%
breaklinks=true,%
bookmarks=true,%
colorlinks=true,%
anchorcolor=black,%
citecolor=blue,
filecolor=black,%
menucolor=black,%
urlcolor=darkblue,%
linkcolor=blue,%
]{hyperref}

\newcommand{\ket}[1]{|#1\rangle}

\usepackage[normalem]{ulem}
\graphicspath{{figs/}}

\begin{document}

\title{Optimal compression of constrained quantum time evolution}

\author{Maurits S. J. Tepaske}
\affiliation{Physikalisches Institut, Universit\"at Bonn, Nussallee 12, 53115 Bonn, Germany}
\author{David J. Luitz}
\affiliation{Physikalisches Institut, Universit\"at Bonn, Nussallee 12, 53115 Bonn, Germany}
\email{david.luitz@uni-bonn.de}
\author{Dominik Hahn}
\affiliation{Max Planck Institute for the Physics of Complex Systems, Noethnitzer Str. 38, 01167 Dresden, Germany}
\date{\today}

\begin{abstract}

The time evolution of quantum many-body systems is one of the most promising applications for near-term quantum computers. However, the utility of current quantum devices is strongly hampered by the proliferation of hardware errors. 
The minimization of the circuit depth for a given quantum algorithm is therefore highly desirable, since shallow circuits generally are less vulnerable to decoherence.
Recently, it was shown that variational circuits are a promising approach to outperform current state-of-the-art methods such as Trotter decomposition, although the optimal choice of parameters is a computationally demanding task. In this work, we demonstrate a simplification of the variational optimization of circuits implementing the time evolution operator of local Hamiltonians by directly encoding constraints of the physical system under consideration. We study the expressibility of such constrained variational circuits for different models and constraints. 
Our results show that the encoding of constraints allows a reduction of optimization cost by more than one order of magnitude and scalability to arbitrary large system sizes, without loosing accuracy in most systems. Furthermore, we discuss the exceptions in locally-constrained systems and provide an explanation by means of an restricted lightcone width after incorporating the constraints into the circuits.

\end{abstract}
\maketitle

\section{Introduction}

Quantum computers are promising computational platforms which are expected to lead to breakthroughs for several difficult problems like integer factorization, optimization or machine learning algorithms by significant quantum speedups over their classical counterparts~\cite{shor_polynomial_1997,kitaev_quantum_1995,ebadi_quantum_2022,Huang2022Quantum}. 
While most tasks will not become feasible before the realization of scalable quantum error correction, the simulation of quantum many-body systems is one of the most promising applications on near term devices~\cite{preskill_quantum_2018}. The non-equilibrium behavior of different condensed matter systems~\cite{bernien2017probing,Gross2017Quantum,hartree_google_2020, scholl2021quantum,ebadi2021quantum, realization_frey_2021, realizing_satzinger_2021, simulating_smith_2019}, lattice gauge theories~\cite{yang2020observation,meth2023simulating} or quantum interactive dynamics~\cite{google2023measurement,noel2022measurement} are examples for applications which were studied recently.  
Current experimental platforms range from tens to more than a hundred qubits~\cite{bluvstein2022quantum,moses2023race}, with thousands of qubits projected for the near future. Such quantum processor sizes becomes increasingly challenging to simulate using the capabilities of classical computers~\cite{arute_quantum_2019, what_zhou_2020, efficient_tindall_2023, classical_anand_2023, kim2023evidence}. 

While the physical system sizes are increasingly impressive, the reachable quantum volume \cite{quantum_moll_2018} of present-day devices is still limited by the presence of noise.
The coupling to the environment, gate imperfections, and measurement errors can destroy the coherence of the quantum state and limit the accuracy of the results of quantum computations. 
There are several attempts to quantify the amount of errors and to find protocols which allow one to average out the errors by increasing the number of required measurements. These techniques can be summarized by the term quantum error mitigation~\cite{scalable_kim_2023, temme_error_2017,guo_quantum_2022,chen_error_2022,filippov_matrix_2022,vovrosh_simple_2021,endo_practical_2018,van_den_berg_model-free_2022,cai_quantum_2022,compressed_tepaske_2023, efficient_li_2017, berg_probabilistic_2022, mitigating_endo_2019, error_kandala_2019}. Recent work has shown that a combination of quantum simulation and quantum error mitigation gives competitive results for the time evolution of quantum systems~\cite{kim2023evidence, classical_anand_2023}.

An alternative approach consists of reducing the circuit depth of the algorithm. The idea is simple: The proliferation of errors is suppressed when the possibilities where such an error can occur are reduced. In other words, one can try to squeeze a given quantum algorithm into a given quantum volume by finding a more efficient circuit representation of the algorithm.
For simulations of time evolution in this era of noisy quantum computing, the state-of-the-art-method is the Trotter decomposition~\cite{theorie_lie_1880,trotter_1959_product,suzuki_1976_generalized,fractal_suzuki_1990}.

Here the unitary operator $U$ generating the time evolution $U\ket{\psi}$ is expressed by a circuit of few-body gates, which can then be evaluated on a quantum computer. The error resulting from this mapping can be controlled by the circuit depth. Trotter circuits have the best known asymptotic error scaling for a large class of local quantum many-body systems and do not require any ancilla qubits \cite{childs_2021_theory}. 
They are projected to be the best iterative algorithms for gate and qubit counts that are expected in the early fault-tolerant era \cite{toward_childs_2018, improved_kivlichan_2020}. Nevertheless, at fixed time the circuit depth can potentially be made smaller by employing numerical optimization.

One promising approach are variational circuits~\cite{the_mcclean_2016, theory_yuan_2019, variational_cerezo_2021, haghshenas2022variational, variational_cristina_2020, local_mizuta_2022, quantum_khatri_2019, noise_sharma_2020, variational_commeau_2020,  quantum_berthusen_2021, real_lin_2021, hardware_benedetti_2021, an_barison_2021, robust_jones_2022, variational_heya_2018, long_gibbs_2022, subspace_heya_2023, Zhao2023Making,zhao2023adaptive, simulating_wada_2022, classically_keever_2023, variational_mansuroglu_2021, Mansuroglu_2023Problem, optimal_tepaske_2023}. The main idea is to describe the time-evolved state or the time-evolution operator using a parametrized circuit. Choosing the parameters boils down to an optimization problem, which can be tackled using the gradient descent method. Several proposed algorithms try to optimize the circuit on a quantum computer or simulator~\cite{robust_jones_2022, an_barison_2021, quantum_berthusen_2021, hardware_benedetti_2021, variational_commeau_2020, noise_sharma_2020, quantum_khatri_2019, local_mizuta_2022, variational_cristina_2020, variational_heya_2018, real_lin_2021, Zhao2023Making, zhao2023adaptive, long_gibbs_2022, subspace_heya_2023, simulating_wada_2022}. However, this is currently unfeasible due to the high error rate on present-day devices.
Other attempts, including our previous work, utilize a circuit optimization algorithm that is specifically designed for classical computers~\cite{Mansuroglu_2023Problem,variational_mansuroglu_2021,optimal_tepaske_2023, classically_keever_2023}. In some cases, we could reduce the circuit depth by almost 50\% in comparison to standard Trotter decomposition at fixed fidelity. While this approach is thus a promising strategy for the future, one major drawback is the large computational cost of the optimization. A possible solution for this issue is to exploit constraints of the system to reduce the number of independent parameters in the circuit. 
These can be encoded directly into the variational circuit ansatz. 

In this work we study this approach for different constraints and analyze the resulting error scaling in comparison to generic variational circuits and Trotter decomposition. 

This paper is organized as follows: In Secs.~\ref{sec:param_circs} and \ref{sec:conserved_charges}, we describe our variational circuits and the implementation of constraints. Furthermore we describe our optimization procedure in detail in Sec.~\ref{sec:optimization}. In Sec.~\ref{sec:results}, we present our results for optimizing the time evolution of three different models with different constraints: The Heisenberg XXZ chain, the PXP model and the quantum link model. Finally, we discuss the implications of our results for encoding constraints into variational circuits in Sec.~\ref{sec:Discussion}.

\section{Methods}\label{sec:method}

In this section, we explain in detail the different circuit architectures that we use to approximate the exact time evolution of a given Hamiltonian. In Sec.~\ref{sec:param_circs}, we introduce a generic brickwall circuit architecture tailored to hardware-native gates. In the following discussions we will refer to these as unconstrained circuits. If the underlying time evolution has conserved charges or local constraints, these can be incorporated into our circuit template. This is described in Sec.~\ref{sec:conserved_charges}. We refer to them as blocked circuits. In Sec.~\ref{sec:optimization}, we describe our cost functions and optimization strategies to determine the free parameters of our circuits. Finally, we explain in Sec.~\ref{subsec:stacking} how the optimized circuits can be used to simulate time evolution on quantum computers for large systems and long time scales beyond the capability of classical devices.

\subsection{Parameterized circuits}\label{sec:param_circs}
\begin{figure}
	\centering
	\includegraphics[width=0.8\columnwidth]{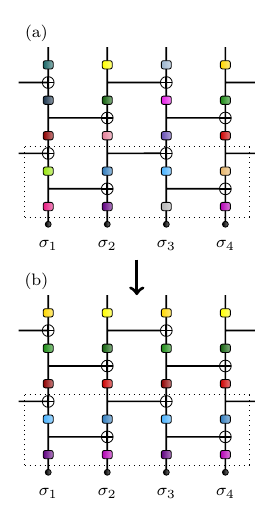}
	\caption{(a) Template for an unconstrained circuit with $M=2$ layers. Each layer of CNOT gates is interspersed with a single-qubit gate layer. At the end of the circuit, we add an extra layer of single-qubit unitaries. Equal colored single-qubit gates are identical. The dotted box encloses one layer of the circuit.
	(b) By adding translation symmetry, we reduce the number of different single-qubit unitaries to two independent single-qubits per half layer. Note that we can not further reduce the number of different unitaries since the brickwall gate layout manifestly breaks site-inversion symmetry. For the rest of the work, we call this architecture translationally invariant variational brickwall~(TIVB) circuits.} 
	\label{fig:bw_circ}
\end{figure}

Since our main intention is to simplify the implementation of the time evolution of a many-body wave function using a quantum computer, it is sensible to use a set of native gates as building blocks. In our case, we use single-qubit gates parametrized as
\begin{equation}
	u(\theta,\phi,\chi) = 
	\begin{pmatrix}
		e^{i\phi}\cos(\theta) & e^{i\chi}\sin(\theta) \\
		-e^{-i\chi}\sin(\theta) & e^{-i\phi}\cos(\theta)
	\end{pmatrix},
	\label{onequbit_param}
\end{equation}
and controlled NOT (CNOT) gates. These gates form a universal gate set~\cite{elementary_barenco_1995}.

The unconstrained circuit architecture $C$ consists of $M$ brickwall layers, as shown in Fig.~\ref{fig:bw_circ}(a). 
Each brickwall layer consists of two half-layers of CNOT gates connecting neighboring qubits, with general single-qubit layers interspersed. At the end of the circuit, we add a final layer of single-qubit unitaries. The advantage of the brickwall architecture is its minimal depth for a fixed amount of CNOTs. This is favorable in the presence of non-correctable errors.

\subsection{Symmetries and constraints}
\label{sec:conserved_charges}

When the targeted unitary $U$ has conserved charges because of an underlying symmetry or local constraint, we can incorporate these into the parameterized circuit by constraining the angles of the one-qubit unitaries or modifying the circuit structure. 

\subsubsection{Lattice symmetries}
\label{sec:lattice_sym}

We are often interested in translationally invariant systems. In this case we choose the one-qubit unitaries as in Fig.~\ref{fig:bw_circ}(b), where we illustrate a translation-invariant circuit with $M=2$ brickwall layers. Here each horizontal layer of one-qubit unitaries contains two independent unitaries (equal colors have equal parameters), giving a circuit with $12M+6$ real parameters.
We note that the translation invariance in a brickwork circuit is necessarily reduced: The minimal unit cell for translations consists of two qubits. This implies that we require two different single-qubit unitaries per half layer. 

Apart from reducing the number of parameters to optimize, the implementation of translation symmetry has additional advantageous implications for the optimization:
A fundamental property that $C$ should reproduce is the Lieb-Robinson bound of correlation spreading for local many-body quantum systems \cite{the_lieb_1972, lieb_bravyi_2006}. 
As a result, if we want to approximate the unitary $U(t)$ for system size $L^*$, and the corresponding lightcone of correlations has width $W(t)<L^*$, then optimizing $C$ at a smaller size $L>W(t)$ is sufficient.

This allows us to perform a difficult optimization procedure for a small amount of qubits and use the result for larger system sizes. We find that this extrapolation works extremely well, sometimes even when $L<W(t)$.
For the timescales we have considered, we find that optimization for $L=8$ qubits is sufficient, with additional optimization at larger sizes giving only insignificant improvements.
For the rest of this work, we have implemented this translation symmetry in our most generic variational circuits, denoting them as translationally invariant variational brickwall~(TIVB) circuits.

\subsubsection{Local constraints}

Apart from lattice symmetries, many systems of interest exhibit symmetric couplings that induce a global symmetry. Here each coupling only connects states with equal global charges. These charges can also be local, as it is the case for gauge theories, or do not need to be associated to a symmetry or gauge freedom at all. Instead, a local constraint is sufficient, as for example in the PXP model \cite{quantum_turner_2018}. 

The constraints or conserved quantities imposes a block-diagonal time evolution operator $U(t)$. Each block corresponds to a different charged sector.
In the following, we want to encode such symmetries or constraints directly into the circuit architecture, which we will then denote as a blocked circuit architecture.
By incorporating such special properties into the circuit architecture, we get a blocked circuit that manifestly has the corresponding block-diagonal structure. This restricts the space of variational circuits to a subspace in which the targeted time-evolution operator $U$ lives, and reduces the parameter count per number of CNOTs. However, it does not guarantee an increased accuracy. In fact, such a restriction can reduce the expressibility, e.g. restricting the maximum possible distance that correlations can travel among the qubits in $M$ layers of CNOTs.

\subsection{Optimization}
\label{sec:optimization}
As a cost function for the optimization of circuits $C$ to faithfully represent a given unitary tranformation $U$, we use the normalized Frobenius distance between $U$ and $C$, 
\begin{equation}
	\epsilon=\frac{||C-U||^2_F}{2^{L+1}} = \frac{\sum_{ij}|C_{ij}-U_{ij}|^2}{2^{L+1}}=1-\frac{\tr[C^{\dagger}U]}{2^L}.
	\label{eq:distance}
\end{equation}
The last equality holds because both $C$ and $U$ are unitary. For small system sizes up to $L\approx16$ this expression can be evaluated exactly, representing $U$ and $C$ as matrices with dimension $2^L$. Otherwise, we represent $U$ and $C$ as matrix-product operators (MPOs), where we inevitably lose information at large $t$ due to entanglement truncation \cite{optimal_tepaske_2023}. To represent the exact evolution operator $U$ as a MPO, we Trotterize the exact time evolution using a sufficiently small step size $\Delta t$ such that the discretization error is smaller than machine precision or truncation errors due to the entanglement barrier.

To determine the parameters of the single-qubit unitaries in the specific architecture, we minimize the distance (\ref{eq:distance}), which we do with first-order gradient descent, using the Adam optimizer \cite{adam_kingma_2014} (see Algorithm~\ref{adam_algo}). This optimizer uses exponentially-decaying averages of the first and second moments of previous parameter updates to modulate the next update.

\begin{figure}
\begin{algorithm}[H]
	\caption{\textit{The Adam optimizer \cite{adam_kingma_2014}.} We want to minimize the distance 
		$\epsilon$ as a function of the the circuit parameters $\vec{\theta}$. Adam uses first-order gradient descent to update the parameters.
		With Adam, we calculate exponentially decaying running averages of the 
		first moment $m$ and the second moment $v$ of the gradient, which are then used to 
		update the parameters as $\delta \vec{\theta} \propto m/\sqrt{v}$. This ensures that 
		the updates are steered away from tiny or huge values to improve convergence to the global minimum, by providing a mechanism to escape local minima without excessively large updates (which are likely to be inaccurate since we are using only local gradient information).}\label{adam_algo}
	\begin{algorithmic}
		\State \textbf{Hyperparameters:}
		\State \quad $\lambda$: Base learning-rate
		\State \quad $\beta_1$: First moment decay rate
		\State \quad $\beta_2$: Second moment decay rate
		\State \quad $\delta$: Regularization
		\State \quad $N_\text{iters}$: Amount of iterations
		\State \textbf{Initial conditions:}
		\State \quad $m_0 \gets 0$ (First moment initially set to zero)
		\State \quad $v_0 \gets 0$ (Second moment initially set to zero)
		\For {( $i=0$; $i < N_\textrm{iters}$; i = i+1 )}
		\State $g_i \gets \nabla_{\vec{\theta}_{i-1}} \epsilon(\vec{\theta}_{i-1})$ (Calculate gradient at current parameters)
		\State $m_i \gets \beta_1 m_{i-1} + (1-\beta_1)g_i$ (Update running average of first moment)
		\State $m^*_i \gets m_i/(1-\beta_1^i)$ (Bias correction)
		\State $v_i \gets \beta_2 v_{i-1} + (1-\beta_2)g_i^2$ (Update running average of second moment)
		\State $v^*_i \gets v_i/(1-\beta_2^i)$ (Bias correction)
		\State $\vec{\theta}_i \gets \vec{\theta}_{i-1}-\lambda m^*_i/(\sqrt{v^*_i} + \delta)$ (Update parameters)
		\EndFor
		\State \Return $\vec{\theta}_i$ (Final circuit parameters)
	\end{algorithmic}
\end{algorithm}
\end{figure}

Our goal is to compress the time-evolution operator $U(t)$ for a sequence of times $t=1,2,...,10$ into parameterized circuits with $M$ layers. The minimization of $\epsilon$ is performed for each pair $(t,M)$ separately. For a fixed layer count $M$, we first optimze the circuit for the smallest timestep $t=1$. We initialize $C$ as the identity circuit and choose a set of Adam hyperparameters $(\lambda,\beta_1,\beta_2,\epsilon_\textrm{reg})$. We perform $\mathcal{O}(10^5)$ iterations of gradient descent to reduce the possibility of getting stuck in local minima. 
For $t>1$, we initialize with the same set of hyperparameters and the optimal result of $t-1$.

We perform this optimization simultaneously for a large grid of hyperparameters. In particular, we choose
\begin{align}
	\begin{split}
		\beta_1,\,\beta_2& \in \{0.9, 0.99, 0.999, 0.9999\}\\
		\epsilon_\textrm{reg}& \in \{10^{-2}, 10^{-4}, 10^{-8}, 10^{-12}\} \\
		\lambda &\in \begin{cases}
			\{0.5, 0.2, 0.1, 0.01, 0.001, 0.0001\},\, \text{blocked}\\
			\{10^{-1}, 10^{-2}, \dots, 10^{-6}\}, \, \text{TIVB circuit}\\
		\end{cases}	
	\end{split}
\end{align}

The difference in the choice of $\lambda$ accounts for the fact that the blocked circuits have less parameters per number of CNOTs. 
After this sequential optimization, we determine the best set of hyperparameters at each $(t,M)$ and use the corresponding parameters to initialize another round of optimization at $L=6$, with the same set of hyperparameters and amount of iterations as before. 
These results are used to initialize the last round of optimization, now for $L=8$, with $\mathcal{O}(10^4)$ iterations. 

Removing any of the sequential optimization steps leads to results that are orders of magnitude worse in accuracy.
Furthermore, we find that using a finer sequential optimization, e.g. starting at $t=10^{-2}$ and progressing in steps of $10^{-2}$, decreases the accuracy of the circuits after every round of optimization.

\subsection{Stacking circuits}\label{subsec:stacking}

The circuits shown in this work are optimized on small system sizes and for short evolution times. As mentioned in Sec. \ref{sec:lattice_sym}, the translation symmetry of the circuit template allows to extend the optimized circuits to arbitrarily large system sizes. 
Furthermore, stacking the circuits~\cite{optimal_tepaske_2023} allows to reach time scales on a quantum device beyond classical capabilities: Consider a circuit, optimized for a given time $t^*$. By repeatedly applying this circuit $n$ times on a quantum computer, this allows to implement the time step $n t^*$ on the quantum device. It was analyzed in Ref.~\cite{optimal_tepaske_2023} that the advantageous scaling in terms of resource cost for the compressed circuit in comparison to Trotter decomposition persists while stacking a circuit multiple times.

\begin{figure}[t]
	\centering
	\includegraphics[width=0.8\columnwidth]{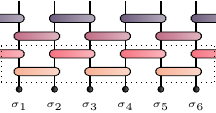}
	\caption{The blocked architecture that we use to compress the XXZ time-evolution operator. Equal-colored gates are identical, with the architecture consisting of U(1)-symmetric two-qubit unitaries (\ref{eq:xxz_param}). The dotted box encloses one layer of circuits. The colors display the manifest translational and bond-inversion symmetries of the XXZ model.} 
	\label{fig:u1_sym_circ}
\end{figure}

\section{Models and results} \label{sec:results}

In this section, we consider specific models and investigate whether shallow depth circuits benefit from incorporating conserved quantities or local constraints into the compressed circuit ansatz $C$.

\subsection{XXZ model}\label{sec:xxz}
We start with the following XXZ model on a periodic chain
\begin{equation}
H=\sum_{\langle i,j\rangle}S^x_iS^x_j + S^y_iS^y_j + \frac{1}{2}S^z_iS^z_j,
\label{eq:ham_xxz}
\end{equation}
which satisfies $[H,Q]=0$ with $Q=\sum_{i=1}^L S^z_i$. The conservation of the total $z$-spin  can be associated with the global $U(1)$ symmetry corresponding to the invariance of (\ref{eq:ham_xxz}) under a simultaneous rotation of all spins in the XY plane. This conserved global charge $Q$ induces a splitting of the time-evolution operator $U(t)=\exp(-itH)$ into $\mathcal{O}(L)$ blocks with fixed $Q$. Specifically, because $Q$ has $L+1$ possible values, i.e. $Q=-L/2,-L/2+1,...,L/2$, there is an equal amount of diagonal blocks. 
The dimension of the largest block, i.e. the zero-magnetization sector, is equal to the binomial coefficient ${L\choose L/2}$. The second-largest sector has ${L\choose L/2-1}$, such that their ratio is $((L/2)!)^2/((L/2+1)!(L/2-1)!)$. This approaches $1$ for large $L$. More broadly, for the XXZ model there is never a single block that significantly outsizes the rest. 

A TIVB circuit architecture from Sec.~\ref{sec:param_circs} does not respect this conservation by default. We can modify this circuit, however, by replacing the elementary CNOT gate with the U(1) symmetric gate
\begin{equation}
U_\textrm{XXZ}(\theta,\phi) = e^{i\theta(\sigma_x\otimes\sigma_x + \sigma_y\otimes\sigma_y) + i\phi\sigma^z\otimes\sigma^z}.
\label{eq:xxz_param}
\end{equation}
Furthermore, we remove the one-qubit unitaries. This is illustrated in Fig.~\ref{fig:u1_sym_circ}.

The gate (\ref{eq:xxz_param}) can be decomposed into three CNOTs with one-qubit unitaries, see e.g. Eq.~(6) in Ref.~\cite{universal_vidal_2004}. This yields a blocked circuit of $\tilde{M}$ U(1)-symmetric brickwall layers with $2\tilde{M}$ parameters. 
The number of CNOT gates is the same as for the previous TIVB circuit with $M=3\tilde{M}$ layers. The lightcone width for a fixed amount of CNOTs is reduced by a factor three, since a minimum of three CNOTs is required to entangle neighboring spins, instead of only one CNOT for the TIVB architecture.

Choosing $\theta=t/2$ and $\phi=t/4$, we obtain the local evolution operator 
\begin{equation}
U_l = e^{-it(S^x_iS^x_j + S^y_iS^y_j + \frac{1}{2}S^z_iS^z_j)},
\end{equation}
so the blocked circuit is a generalization of first-order Trotter decomposition with variational time steps \cite{quantum_berthusen_2021, variational_mansuroglu_2021,Zhao2023Making, zhao2023adaptive}. We find that compressing into the second-order Trotter layout, i.e. adding an extra half-layer on top of the circuit in Fig.~\ref{fig:u1_sym_circ}, provides no advantage over the first-order layout. To be more concrete, the optimized $\epsilon$ decreases smoothly as we add half-layers to the architecture. We checked that in both cases we get the Trotter scaling from Ref.~\cite{first_layden_2022} when stacking the optimized circuits. 

In order to see the advantages of the blocked architecture, we benchmark it against the TIVB architecture. We compare blocked circuits with $\tilde{M}=1,3,4,5,7,8$ layers against TIVB circuits with $M=4,8,12,16,20,24$ layers.
As we discussed before, each $U(1)$-conserving two-qubit gate can be expressed by a combination of unitaries and at most three CNOT layers. The number of CNOT gates for a blocked circuit with $\tilde{M}$ layers is thus the same as for the previous TIVB circuit with $M=3\tilde{M}$ layers. Therefore we have a slight mismatch in our comparison by means of circuit depth, since only for $\tilde{M}=4,8$ can we exactly match the CNOT count with $M=12,\,24$. The other compared circuits differ at most by one additional brickwall layer of CNOTs.

\begin{figure}
        \centering
	\includegraphics[width=1.0\columnwidth]{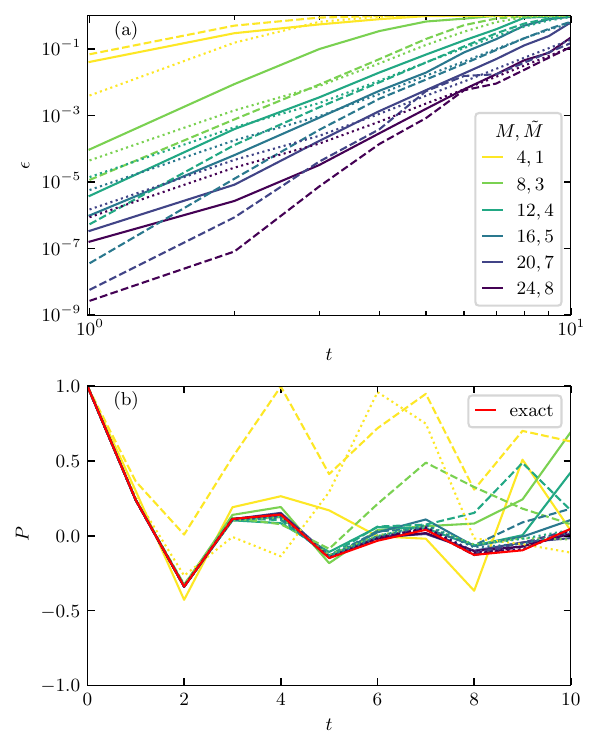}
\caption{Results for the compression of the XXZ time-evolution operator for system size $L=16$ and times up to $t=10$ for brickwall circuits for a different number of layers $M$. We compare TIVB circuits with $M=4,8,12,16,20,24$ with blocked circuits with $\tilde{M}=1,3,4,5,7,8$. 
\textbf{(a)} The normalized distance $\epsilon$ of the optimized circuits as a function of the evolution time $t$, for the TIVB circuits~(solid lines), blocked circuits~(dashed lines) and second-order Trotter decomposition with $\tilde{M}+1/2$ steps~(dotted lines). The blocked architecture outperforms the TIVB architecture in $\epsilon$ per number of CNOTs for all $t$. Both parameterized architectures outperform the second-order Trotter circuits.
\textbf{(b)} The $z$-magnetization imbalance $P$ of the propagated N\'eel state $\ket{\mathbb{Z}_2}$ for the different circuits architectures. As before, the blocked architecture outperforms the other approaches, as long as $\epsilon$ is on the order of $1\%$ at most. The observable is calculated on a time grid with spacing $\delta t=1$ and lines are guides to the eye.}\label{fig:1_xxz}
\end{figure}

We minimize Eq.~(\ref{eq:distance}) for the time-evolution operator of (\ref{eq:ham_xxz}) for times $t=1,2,...,10$. The results are shown in Fig.~\ref{fig:1_xxz} for system size $L=16$. We optimized the parameters of the circuit for system sizes $L=6$ and $L=8$ and interpolated to larger system sizes by imposing translation symmetry as described above. We compare the error for TIVB~(solid lines), blocked~(dashed), and second-order Trotter circuits with $\tilde{M}+1/2$ layers~(dotted).

In Fig.~\ref{fig:1_xxz}(a), we show the normalized distance $\epsilon$ of the optimized circuits. The blocked circuits outperform the TIVB circuits for all investigated timesteps $t$. This difference is most prominent at short times and vanishes for large times. For small timesteps $t$, the second-order Trotter circuits perform worse than the optimized circuits. For large $t$, the performance is similar to the optimized blocked circuit.
Even if this observation is tight only for $M=12,24$ due to the equal number of CNOT gates, it extends to other numbers of layers $M$. However, in this case the comparison is more difficult due to the mismatch in CNOTs. 

We are not only interested in the distance between the circuit and the targeted unitary, but also in the resulting error for measurable observables. Moreover, we want to know if optimizing the full distance also yields systematically-increasing accuracy on the dynamics of the biggest block. As a check for that, we show in Fig.~\ref{fig:1_xxz}(b) the $z$-magnetization imbalance
\begin{align}\label{eq:imbalance} 
P=\sum_j(-1)^j\langle \sigma^z_j\rangle
\end{align}
for the N\'eel-state 
\begin{align}\label{eq:Neelstate}
\ket{\mathbb{Z}_2}=\ket{\uparrow\downarrow\uparrow...}
\end{align}
 propagated to time $t$ by the different circuit architectures. The time evolution of $\ket{\mathbb{Z}_2}$ is constrained to the largest block when the global U(1) constraint is satisfied. We compare with the exact time evolution obtained from exact diagonalization~(red line). The blocked circuits reproduce $P$ with the highest accuracy as long as at least $\epsilon\sim\mathcal{O}(10^{-2})$. As an example, for $M=12$ the blocked circuit is best until $t=5$, where its error is a few percent. The blocked circuit with $M=24$ is best until $t=8$, after which the Trotter circuit is best. For larger distance $\epsilon$, the TIVB and Trotter circuits perform better, with almost all Trotter curves lying around the exact curve until $t=10$. As long as at most $\epsilon\sim\mathcal{O}(10^{-2})$, the accuracy on $P$ increases systematically, i.e. determining a deeper circuit with lower $\epsilon$ generally also improves the accuracy on $P$.

It is unclear whether this situation will change significantly when performing a simulation on noisy quantum computers. In that case, we can perform classical post-processing based on the constraint, discarding output states that violate it, in an effort to mitigate errors \cite{chen_error_2022}. Furthermore, it should be noted that the TIVB architecture has more than eighteen times as many parameters as the blocked architecture for an equivalent circuit depth. This makes the backpropagation computations during the optimization process at least eighteen times as expensive.

In App.~\ref{app:restricted_distances} we show that the relative performance of the blocked and TIVB circuits cannot be improved by constraining the cost function to one of the largest blocks and its off-diagonal.

\subsection{PXP model}\label{subsec:PXP}

The PXP model on a chain of $L$ qubits with periodic boundary conditions is given by \cite{quantum_turner_2018}
\begin{equation}
	H=\sum_{j=1}^L P^z_{j-1}\sigma^x_jP^z_{j+1},
\label{eq:PXP_ham}
\end{equation}
where $\sigma^\alpha_j$ denote the Pauli operators acting on spin $j$ and $P^z_{j}=(1-\sigma^z_j)/2$ is a projector of the $j$-th qubit into its ground state. The three-body term introduces a local constraint: neighboring excited states are immobile and cannot be created or annihilated. This can be associated with an extensive number of conserved local charges $Q_j=(1+\sigma^z_j)(1+\sigma^z_{j+1})$ that encode the absence ($Q_j=0$) or existence ($Q_j=1$) of a frozen pair on the bond between sites $j$ and $j+1$. The time-evolution operator splits into $\mathcal{O}(2^L)$ blocks, each block being labeled by a particular charge configuration $\{Q_j\}$. The largest block has $Q_j=0\ \forall j$ and is well known for its weak ergodicity breaking: its Hilbert space has an exponentially small fraction of ETH violating states, as revealed by the revivals in return probability and $z$-magnetization imbalance that occur when time-evolving N\'eel product states \cite{quantum_turner_2018}. 

Besides the PXP model having exponentially many more blocks than the XXZ model (\ref{eq:ham_xxz}), another qualitative difference is the relative size between the largest and second largest blocks: The largest block has dimension $F_{L-1}+F_{L+1}$, where $F_{j}$ is the Fibonacci sequence $F_0=0,F_1=1,...$ \cite{quantum_turner_2018}. The second-largest block has a dimension equal to that of the largest block of the $L-4$ PXP model with OBC, i.e. $F_{L-2}$. Their ratio approaches $5.8541...$ for $L\rightarrow\infty$. The PXP model thus has a dominant block in the Hilbert space, in contrast to the XXZ model where this ratio approaches $1$.

\begin{figure}[t]
        \centering
        \includegraphics[width=1.0\columnwidth]{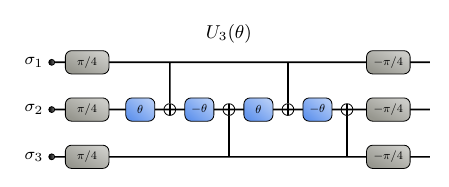}
\caption{The three-qubit circuit $U_3(\theta)$ that implements a PXP-blocked rotation on the middle site. The blue unitaries correspond to $u(\theta,0,0)$ and the gray to $u(0,\phi,0)$. The gray and CNOT gates ensure that $u(4\theta,0,0)$ is applied to the inner qubit only when its neighboring qubits are in the ground state.}
\label{fig:block_gate}
\end{figure}

\begin{figure*}
        \centering
	\includegraphics[width=1.0\textwidth]{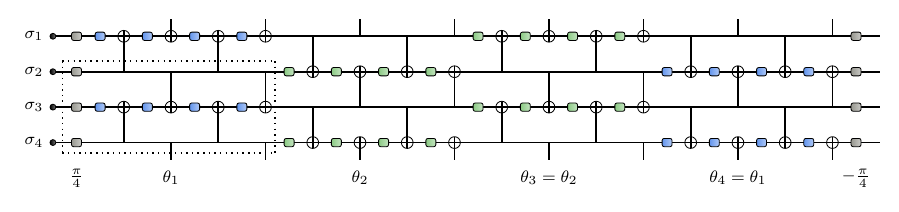}
\caption{The blocked architecture with $M=8$ brickwall layers. The building blocks is the sub-circuit $U_3(\theta)$ defined in Fig.~\ref{fig:block_gate} and indicated by a black dashed box. The implemented circuit has a time-inversion symmetry, resulting in $M/4$ parameters for $M$ brickwall layers. The gray unitaries correspond to $u(0,\pm\pi/4,0)$, and the blue and green unitaries to $u(\theta_j,0,0)$, with equal-colored gates being equivalent.}
\label{fig:block_circuit}
\end{figure*}

\subsubsection{Circuit architectures}
We compare again the compression of the PXP time-evolution operator into TIVB and blocked circuit architectures. The TIVB architecture is the same as in Fig.~\ref{fig:bw_circ}, which breaks the site-inversion symmetry of the PXP model (\ref{eq:PXP_ham}). 

In order to incorporate the local constraint of the PXP model, we have to fix the relation between different one-qubit angles. To do so, we consider a blocked circuit $U_3(\theta)$ acting on three qubits that satisfies the local constraint. It has a single parameter and is displayed in Fig.~\ref{fig:block_gate}.  
In this figure, the blue unitaries denote $u(\theta,0,0)=\exp(i\theta\sigma^y)$. The gray unitary gates correspond to $u(0,\pm\pi/4,0)$. The interplay of $u(0,\pm\pi/4,0)$ and the CNOT gates ensures that the inner qubit is only rotated when the neighboring qubits are in the ground state. 
With the choice $\theta=t/4$, it is an implementation of the local time-evolution operator $U_j=\exp(-itP_{j-1}\sigma^x_jP_{j+1})$.

It is important to note that we can generalize $U_3(\theta)$ to a full circuit which implements the constraint on all even qubits (and then on all odd qubits) simultaneously. This allows us to construct a blocked circuit as shown in Fig.~\ref{fig:block_circuit}, implementing the PXP constraint for the entire system. From this circuit diagram it is immediately clear that information can only travel one site after four brickwall layers of CNOT gates. 

We find that we can implement a time-inversion symmetry without loosing accuracy, i.e choosing the angles $\theta_i$ symmetric as $\theta_i=\theta_{M/2-i}$. The resulting brickwall circuit with $M$ layers has $M/4$ free parameters.

Similarly to the U(1)-symmetric circuits in Sec.~\ref{sec:xxz}, the blocked ansatz is a first-order Trotter circuit with variational time steps, and it always satisfies ${M \text{ mod }4=0}$ due to its construction. It gives the identity operator when all angles are set to zero. 

\begin{figure}
        \centering
	\includegraphics[width=1.0\columnwidth]{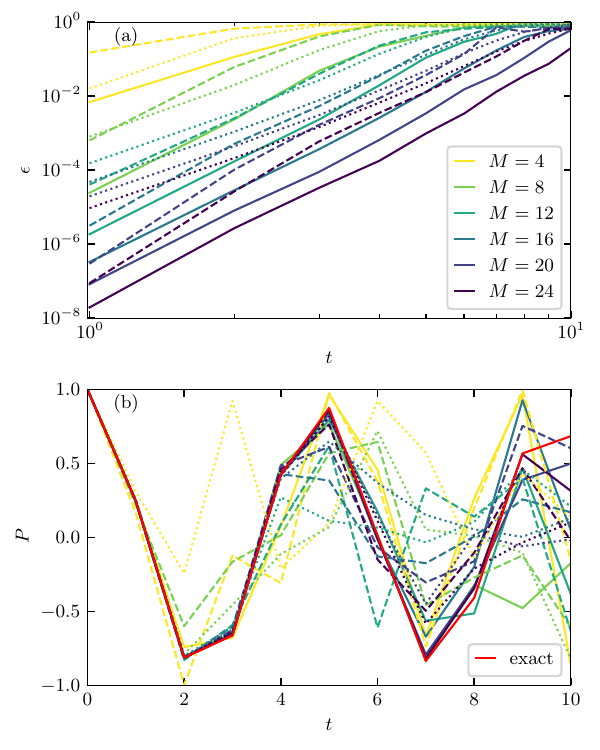}
\caption{Results for the compression of the PXP time-evolution operator for system size $L=16$ and times up to $t=10$ for brickwall circuits with up to $M=24$ layers. \textbf{(a)} The normalized distance $\epsilon$ as a function of the timestep $t$, results for the TIVB circuits~(solid lines), blocked circuits~(dashed) and second-order Trotter circuits with $M+2$ layers~(dotted). The TIVB architecture outperforms the blocked architecture with the same circuit depth. Both architectures outperform the second-order Trotter circuits, although the blocked architecture is only marginally better at large $t$. \textbf{(b)} The imbalance $P$ (\ref{eq:imbalance}) as a function of time, starting from the N\'eel state. The exact result is shown as a solid red line. The same qualitative picture emerges as in panel (a).}\label{fig:1}
\end{figure}

We compress the PXP time-evolution operator for times $t=1,2,...,10$ into TIVB and blocked circuits with $M=4,8,...,24$ brickwall layers of CNOTs. The results are shown in Fig.~\ref{fig:1}, evaluated at $L=16$. The TIVB circuits are shown as solid lines, the blocked circuits as dashed lines, and the second-order Trotter circuits with $M+2$ layers as dotted lines.

\subsubsection{Distance and imbalance}
\label{sec:distance_and_imbalance}

In Fig.~\ref{fig:1}(a) we show the normalized distance $\epsilon$. The accuracy of the TIVB circuits is at least one order of magnitude better than the blocked circuits for fixed circuit depth. This is in contrast to the case for the XXZ chain. As before, both architectures outperform the Trotter circuits in terms of distance.  

We evolve the N\'eel state (\ref{eq:Neelstate}) up to a time $t$ and measure the imbalance $P$ defined in Eq.~(\ref{eq:imbalance}). 
In the case of the PXP model, the dynamics gives rise to prominent revivals in the imbalance $P$, since the initial $\ket{\mathbb{Z}_2}$ state has large overlap with scarred eigenstates of the model~\cite{quantum_turner_2018, chen_error_2022}. 
In Fig.~\ref{fig:1}(b) we show the results for the optimized circuits from Fig.~\ref{fig:1}(a). The red line displays the exact values $P_\textrm{exact}$ obtained from exact diagonalization. 

Although we focus here on a state that evolves within a single block of the Hilbert space, the same picture as in Fig.~\ref{fig:1}(a) carries over: The TIVB circuits outperform the blocked circuits with the same circuit depth. For a fixed amount of gates and a fixed accuracy threshold $\epsilon$, the TIVB circuits can reach times around $2t$ when the blocked and Trotter circuits reach $t$. Similar to the situation for the XXZ model, the blocked circuits outperform the Trotter circuits when $\epsilon$ is of order $\mathcal{O}(10^{-2})$ at most. Afterwards, there is a short region where the Trotter circuit is most accurate, before both the Trotter and constrained circuits are inaccurate. 
This is in contrast to our results for the XXZ model, where the Trotter circuits remained fairly accurate for the magnetization even when $\epsilon$ was highly inaccurate.

\begin{figure}
        \centering
	\includegraphics[width=1.\columnwidth]{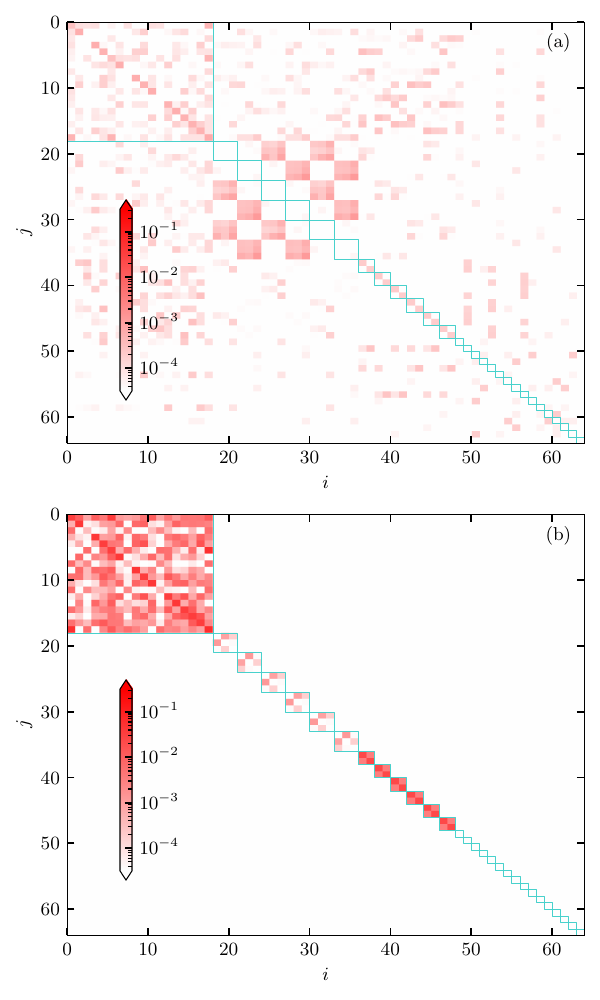}
\caption{The component-wise absolute error $|C_{ij}-U_{ij}|^2$ of a TIVB circuit (a) and a blocked circuit (b) with $M=20$ brickwall layers for $L=6$ at $t=5$. The blocks are shown as turquoise boxes. The TIVB circuit connects different subspaces: It hybridizes blocks that have a fixed amount of frozen pairs but at slightly shifted locations. This hybridization is accompanied with a higher approximation accuracy inside of the blocks, where the errors are more than an order of magnitude smaller than for the blocked circuit architecture with the same circuit depth.}
\label{fig:1split2}
\end{figure}

\subsubsection{Detailed error analysis}
To understand the origin of the discrepancy between the architectures, we consider the component-wise absolute error $|U_{ij} - C_{ij}|^2$ for $L=6$ at time $t=5$ with $M=20$ in Fig.~\ref{fig:1split2}. 
We compare the TIVB architecture~(left), with a total distance $\epsilon\approx10^{-2}$, with the blocked architecture~(right), which is significantly worse with $\epsilon\approx10^{-1}$.
The block structure of the Hilbert space is indicated by turquoise lines. The main contributions of errors in the TIVB circuit stem from violation of the block structure.
In particular, we find that the optimization systematically leads to hybridization of blocks that have an equal amount of frozen spin pairs, i.e it allows for small moves of the frozen spin-up pairs. 

As a compensation, this allows a significant increase of expressibility within the diagonal blocks in comparison to the blocked architecture: The errors within a diagonal block are more than one order of magnitude lower on average. In case the violation of the constraint is not severe, this still allows the use of post-processing quantum simulation data in real experiments, e.g. discarding measurement output which violates the constraint. This improves the mitigation of errors in a quantum simulation.
The downside of the TIVB architecture is the optimization cost: This architecture has $48$ times more parameters than the blocked circuits for fixed circuit depth. This increases the cost of backpropagation during the optimization process by at least the same amount.

\begin{figure}
	\centering
	\includegraphics[width=1.\columnwidth]{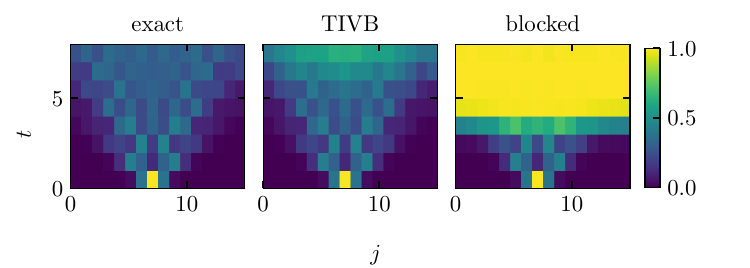}
	\caption{The out-of-time-ordered correlators $C_{L/2,j}(t)$ (\ref{eq:Otoc}) as a function of time $t$ and position $j$, traced over the largest subspace. Results are shown for the exact $L=16$ PXP time-evolution operator (left panel), the TIVB brickwall circuit approximation (center), and the blocked brickwall circuit approximation (right). Both circuits have a fixed depth of $M=16$ layers for all times $t$. The architecture of the blocked circuit constraints its correlation range. As a consequence, it can reproduce $C_{L/2,j}(t)$ only for times $t<4$. In contrast, the TIVB architecture is only limited by the expressibility due to its given circuit depth.}\label{fig:2}
\end{figure}

\subsubsection{Growth of correlations}
To gain more insight, we analyze the results for the out-of-time-ordered correlator (OTOC)
\begin{align}\label{eq:Otoc}
	C_{L/2,j}(t)=||[\sigma^z_{L/2}(t),\sigma^z_j]||^2_F 
\end{align}
as a function of $j$ and $t$. The average is taken only over the largest block without frozen spin-up pairs.  
The results are shown in Fig.~\ref{fig:2} for a system with $L=16$ sites, up to time $t=8$. In the left panel we show the exact results. In the middle and right panels, we look to reproduce this using the TIVB and blocked architecture with $M=16$ layers.

Due to the specific construction of $U_3$ shown in Figs.~\ref{fig:block_gate} and \ref{fig:block_circuit}, the blocked circuit can only reproduce correlations within $M/4$ sites. As a result, the blocked architecture has a reduced maximum-velocity lightcone width, which limits its expressibility. As is shown in Fig.~\ref{fig:2}, this restricts the reachable times of the blocked circuit: In case of $M=16$, the maximum reachable time is thus $t=4$. In contrast, the TIVB architecture has no such limitation and is capable to reproduce the lightcone structure of the OTOC up to $t=8$. 

\subsubsection{Scaling with system size}
\begin{figure}
        \centering
	\includegraphics[width=1.0\columnwidth]{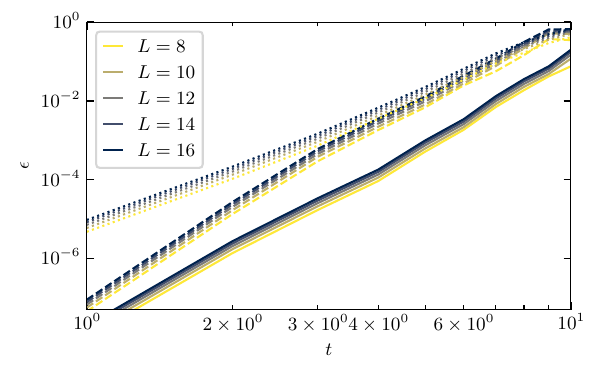}
\caption{The system-size extrapolation of the normalized distance $\epsilon$ for the TIVB (solid), blocked (dashed) and second-order Trotter circuits with $M=24$ layers (dotted) at times $t=1,2,...,10$. The circuits are optimized at $L=8$ and then used to evaluate the quantities at larger sizes $L=10,12,14,16$ by exploiting translational invariance. All optimized circuits extrapolate as good as the Trotter circuits, with errors that show only weak dependence on system size. Thus the optimization of circuits is scalable to larger system sizes without losing accuracy.}
\label{fig:3}
\end{figure}

Now we will check the scaling of accuracy with system size. We use the $M=24$ circuits optimized at $t=1,2,...,10$ with $L=8$ to calculate their distances to the exact time-evolution operator at larger sizes $L=10,12,14,16$. In Fig.~\ref{fig:3} we present the results. All architectures have a weak dependence on system size $L$. This shows the scalability of our approach to system sizes larger than investigated in this work.

\begin{figure}
        \centering
	\includegraphics[width=1.\columnwidth]{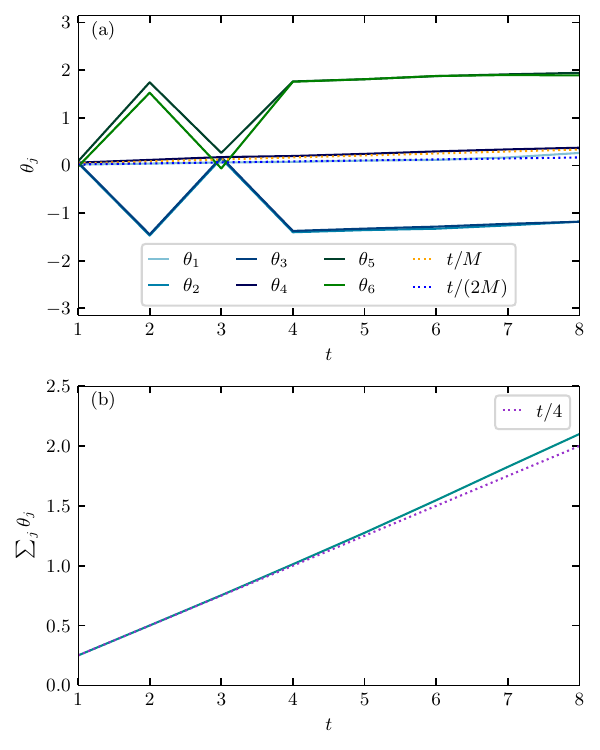}
\caption{\textbf{(a)} The six optimized angles $\theta_l$ of the blocked architecture with $M=24$ brickwall layers. The angles of the second-order Trotter circuit with two extra brickwall layers are shown as dotted lines. While the angles  of the Trotter circuit $\theta_l$ are positive, the optimized blocked circuit relies on the cancellation of multiple forward and backward evolutions. \textbf{(b)} The sum of the optimized angles from panel (a) shown as a solid line, and the sum of the Trotter angles $t/4$ shown as a dotted line. Both angles sum up to the same value of $t/4$, as long as the blocked circuit has an optimized distance below $\epsilon\approx10^{-1}$. The sequences of forward and backward evolutions observed in panel (a) always accumulate to $t/4$ when it is sufficiently accurate.}
\label{fig:6}
\end{figure}

\subsubsection{Comparison with Trotter decomposition}
In order to compare the optimized blocked circuits with the Trotter circuits, we analyze the optimized circuit parameters $\theta_l$. As mentioned above, the Trotter circuits have the same structure as the blocked circuits, but with fixed angles $\theta_1=t/(2M)$ and $\theta_l=t/M$ for $l>1$.

In Fig.~\ref{fig:6}(a), we compare the six angles $\theta_l$ of the blocked circuit with $M=24$ layers at various times and compare it with the angles of the Trotter circuit. In Fig.~\ref{fig:6}(b), we compare the sum of the optimized angles $\sum_j\theta_j$ as a blue solid line, with the sum of Trotter angles $t/4$ shown as a dotted line. 

The optimized angles are far off the Trotter parameters in most cases. The blocked circuits implement a sequence of forward and backward time evolutions. In contrast their sums $\sum_j\theta_j$ are close as long as $\epsilon\sim\mathcal{O}(10^{-2})$. Higher-order Trotter circuits also contain backward time evolution \cite{fractal_suzuki_1990}, but we did not find any symmetrical Trotter decomposition that matches our optimized angles. Specifically, the angles of the often-used decompositions from Ref. \cite{fractal_suzuki_1990} are significantly smaller. 

In App.~\ref{app:pxp_detail} we consider the scar states of the PXP model in more detail. In App.~\ref{app:restricted_distances} we consider the distance (\ref{eq:distance}) restricted to various blocks. There we also show that the distance of the largest block is not improved by restricting the cost function to it. We show this for the PXP and XXZ models.

\subsection{Quantum link model}\label{subsec:qlm}

\begin{figure}
        \centering
	\includegraphics[width=1.0\columnwidth]{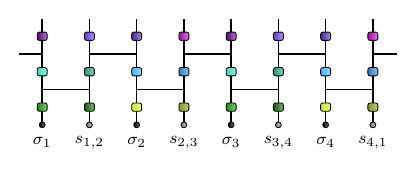}
\caption{One layer of the TIVB architecture used for the quantum link model (\ref{eq:ham_link}). Each horizontal layer of one-qubit unitaries contains four distinct one-qubit unitaries.}\label{fig:link_circsfree}
\end{figure}

\begin{figure}
        \centering
	\includegraphics[width=1.\columnwidth]{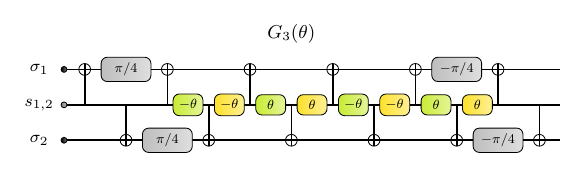}
\caption{CNOT implementation of the parameterized constrained coupling $G_3(\theta)$ between a gauge field and its neighboring matter sites. It contains a single parameter $\theta$. The gray one-qubit unitaries are $u(\pm\pi/4,0,0)$, the green are $u(\pm\theta,-\pi/2,0)$ and the yellow are $u(\pm\theta,\pi/2,0)$.}\label{fig:link_gate}
\end{figure}

As a third model, we consider the massless spin-$1/2$ quantum link model (QLM)
\begin{equation}
H=-\sum_{j=1}^L \sigma^+_js^+_{j,j+1}\sigma^-_{j+1} + \textrm{h.c.},
\label{eq:ham_link}
\end{equation}
which describes matter spins $\sigma_j$ that are coupled by gauge spins $s_{j,j+1}$ that live on the links. It is equivalent to a discrete massless Schwinger model with staggered fermions~\cite{quantum_hauke_2013}. The time-evolution operator splits into exponentially many blocks, to account for the gauge freedom that is generated by the conserved charges
\begin{equation}
Q_j=(\sigma_j^z+s^z_{j-1,j}-s^z_{j,j+1}+(-1)^j)/2,
\label{eq:gauge_charge}
\end{equation} 
which act on a matter spin and its neighboring gauge spins. It takes on the values $Q_j=0,\pm1,\pm2$. 

One of the two largest blocks corresponds to the gauge-invariant sector $Q_j\ket{\psi}=0$ $\forall j$. When tracing out the matter spins in this sector, we get a PXP model for the gauge spins \cite{lattice_surace_2020}. As such, the gauge-invariant block is simply the largest block of a PXP model with $L$ sites. The other largest block corresponds to the charged sector with $Q_j\ket{\psi}=(-1)^j\ket{\psi}$, which is the largest block of a dual PXP model where neighboring gauge spins are frozen when they both point down. The next-largest blocks correspond to the charge configurations $Q_j$ that can be obtained from the gauge-invariant sector by replacing pairs of neighboring $Q_j=0$ with $Q_j=(-1)^j$, leaving at least one pair of zeros. The ratio of the dimensions of the largest and second-largest blocks is $1.206...$. Hence there are many similarly-sized blocks, unlike the PXP model where the largest block is multiple times larger than the rest. The situation is analogous to the XXZ model, where at finite $L$ the second-largest block is only marginally smaller than the largest block. However, we now have $L$ conserved charges instead of one, such that there are exponentially more blocks (that are therefore exponentially smaller).

\begin{figure}
        \centering
	\includegraphics[width=0.8\columnwidth]{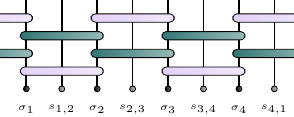}
\caption{The translationally-invariant blocked architecture with $M=2$ layers, where every three-qubit ``gate'' is a $G_3(\theta)$ subcircuit. The circuit has a time-inversion symmetry, as indicated by the colors, such that a blocked circuit with $\tilde{M}$ layers has only $\tilde{M}$ parameters.}\label{fig:link_circs}
\end{figure}

To account for the doubled unit cell of the QLM, we compress its time-evolution operator into TIVB brickwall circuits that contain four unique one-qubit unitaries per half-brickwall layer. This is illustrated in Fig.~\ref{fig:link_circsfree} for $M=1$. This TIVB architecture has $24M+12$ parameters. Due to the chosen layout of matter and gauge spins, each elementary nearest-neighbor gate always acts on a matter and gauge spin.  
To compress while respecting the local constraint, we start by decomposing the local time-evolution operator into nearest-neighbor CNOTs and one-qubit unitaries. This yields the circuit $G_3(\theta)$ shown in Fig.~\ref{fig:link_gate}. The local time-evolution operator is obtained by setting $\theta=t/8$. 
$G_3(\theta)$ is our building block for a charge-conserving circuit, as is shown in Fig.~\ref{fig:link_circs} for $\tilde{M}=2$ layers. Each three-qubit gate represents a two-local subcircuit $G_3(\theta)$, and equal-colored gates have equal angles in order to enforce translation and time-inversion symmetry. Consequently, the blocked circuit with $\tilde{M}$ layers has only $\tilde{M}$ parameters, and has an equal amount of CNOTs as a brickwall circuit with $M=6\tilde{M}$. However, because a blocked circuit cannot be cast into a two-local brickwall circuit, the comparison at fixed gate count is not entirely straightforward.

\begin{figure}[t]
	\centering
	\includegraphics[width=1.0\columnwidth]{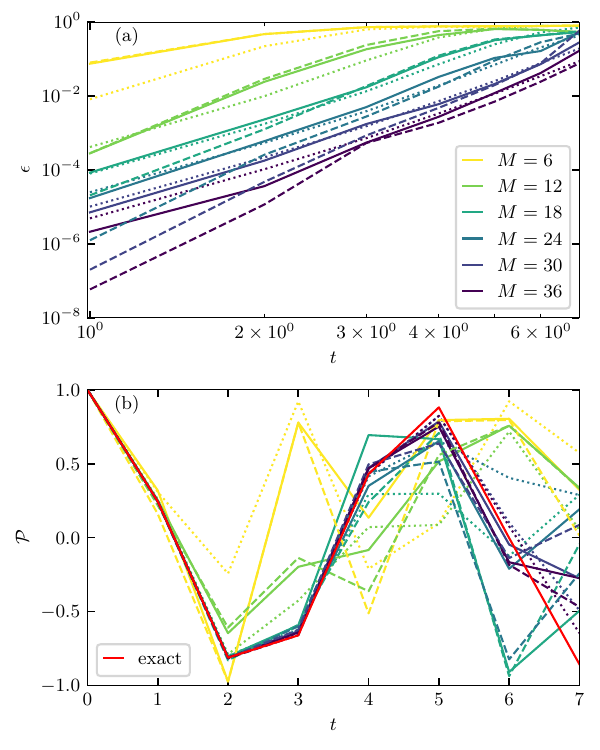}
	\caption{Results for the compression of the massless spin-1/2 quantum link model time-evolution operator for system size $L=16$, counting both matter and gauge spins, and times up to $t=7$ for brickwall circuits with up to $M=36$ layers. The blocked circuit with $\tilde{M}$ layers has an equal amount of CNOTs as the TIVB circuit with $M=6\tilde{M}$, but the CNOT architecture is deeper in the blocked case. For convenience we label $\tilde{M}$ as $6M$ in the plots. \textbf{(a)} The normalized distance $\epsilon$ of the optimized circuits. We compare TIVB circuits~(solid lines), blocked circuits~(dashed) and second-order Trotter circuits with $\tilde{M}+1/2$ layers~(dotted). The blocked circuits outperform the others for fixed gate count, most prominently at low $t$. \textbf{(b)} The string-order parameter $\mathcal{P}$ of the gauge spins, evaluated for the circuits from panel (a). The exact result is shown as a red line. As for the PXP model, this quantity shows revivals. The relative performance observed in panel (a) carries over for the circuits with $\epsilon$ below some threshold of order $\mathcal{O}(10^{-2})$.}\label{fig:1_link}
\end{figure}

The performance of the circuit compressions for the QLM model is shown in Fig.~\ref{fig:1_link}. 
The normalized distance $\epsilon$ for the different circuit architectures is shown in Fig.~\ref{fig:1_link}(a). The situation is analogous to that for the XXZ model from Fig.~\ref{fig:1_xxz}. The blocked circuits outperform the other approaches for fixed gate count when $\epsilon\leq10^{-2}$, and afterwards they all perform similarly. For the circuit sizes that we have considered, any significant advantage over the Trotter circuits is restricted to $t<3$.

The curves for $M=6,12$ are an exception, since here the second-order Trotter circuits appear to outperform the optimized circuits. However, the Trotter architecture gets an extra half-layer in Fig.~\ref{fig:link_circs}, such that in Fig.~\ref{fig:1_link} we e.g. compare a blocked circuit with $\tilde{M}=1$ with a Trotter circuit with $\tilde{M}=1+1/2$. The difference is three brickwall layers worth of CNOT gates. This difference becomes negligible at large $\tilde{M}$, but for small $\tilde{M}$ it skews the comparison, as we see for $M=6,12$ in Fig.~\ref{fig:1_link}.

As a further test, we simulate the dynamics of the gauge-invariant block, again in order to determine whether optimizing the global distance $\epsilon$ also yields systematic improvement on a block-restricted quantity. Specifically, because the QLM restricted to the gauge-invariant sector is a PXP model on the gauge spins, the propagation of states in this block can again yield revivals \cite{lattice_surace_2020}. For this correspondence, the gauge $z$-spin operator is mapped onto a staggered $z$-spin operator. Consequently, revivals now occur for uniform up or down states of the gauge spins, as captured by the magnetization
\begin{equation}
	\mathcal{P}=\sum_{i=1}^L\langle s^z_{i,i+1}\rangle/L.
\end{equation}
We consider the initial state
\begin{align}
 \ket{\mathbb{Z}^{\text{QLM}}_2}=\ket{\downarrow_{\sigma_1}\downarrow_{s_1}\uparrow_{\sigma_2}\downarrow_{s_2}\downarrow_{\sigma_3}\downarrow_{s_3}...},
\label{eq:z2_qlm}
\end{align}
which is the product of a staggered state for the matter spins $\ket{\downarrow_{\sigma_1}\uparrow_{\sigma_2}...}$ and a uniform down state for the gauge spins $\ket{\downarrow_{s1}\downarrow_{s2}...}$. The state (\ref{eq:z2_qlm}) is part of the gauge-invariant sector, in which it remains as long as it is evolved with a circuit that obeys the local constraint. We show the results in Fig.~\ref{fig:1_link}(b), where the exact curve is shown as a red line. As for the other models, the picture from panel (a) carries over for the circuits with distance smaller than $\epsilon \lesssim\mathcal{O}(10^{-2})$. For larger compression errors, all circuits show large deviations from the exact curve. 

\section{Discussion}\label{sec:Discussion}

In this work, we have studied the performance of different circuit architectures to simulate time evolution in the presence of conserved charges.
In particular, we put an emphasis on the accuracy that the compressed circuits can achieve when we directly encode constraints into the circuit architecture.

In the presence of global charges and also for lattice gauge theory, we can decrease the optimization cost by more than two orders of magnitude and simultaneously increase the accuracy. At the same time, translation invariance allows for a scalability of our circuits to arbitrary large system sizes while remaining accurate.
The only exception arises for systems with local constraints such as the PXP model. In this case, the expressibility of the blocked circuits is reduced to the extent that lifting the constraint gives rise to substantial improvement. It manifests itself as a severely restricted lightcone of correlation spreading~(see Fig.~\ref{fig:2}) and thus imposes fundamental limitations on the accuracy that can be reached with shallow circuits.

Furthermore, we believe that the ability of the TIVB architecture to outperform the blocked architecture on only one of the three considered models is related to the different symmetry block structures of the Hamiltonians.
As shown in Fig.~\ref{fig:1split2}, lifting the constraint allows for a hybridization of different symmetry blocks in the PXP model. While this induces additional errors violating the constraint, the hybridization of different blocks allows a larger expressibility and thus error reduction within a symmetry block.

Another interesting observation arises for the blocked circuits: In many cases, the constraint puts such heavy constraints on the final architecture that they automatically turn out as a Trotter circuit with variational timesteps. However, our optimized circuits did not coincide with any conventional Trotter decompositions, relying on the cancellation of relatively large forward and backward local time evolutions. In the case of the PXP model they achieved an increase in accuracy of more than an order of magnitude in comparison to standard second order Trotter decomposition with the same gate count, as measured in the normalized distance. It is desirable to get a more rigorous understanding of the origins of this improved accuracy~\cite{childs_2021_theory}.

Another interesting direction for future studies is the use of different cost functions to simulate quantum dynamics. In the case of adaptive Trotter decomposition, it was shown that bounding the errors in the mean and the variance of the energy is sufficient to obtain very precise expectation values for local observables at intermediate to long times~\cite{zhao2023adaptive,Zhao2023Making}. This raises the question whether we can use shallower circuits if we only care about reproducing the time evolution of local observables.

\begin{acknowledgments}
We thank Zlatko Minev, Alireza Seif, Oles Shtanko and Derek Wang for inspiring discussions.
This work was supported by the Deutsche Forschungsgemeinschaft through the cluster of excellence ML4Q (EXC2004, project-id 390534769). We also acknowledge support from the QuantERA II Programme, which has received funding from the European Union’s Horizon 2020 research innovation programme (GA 101017733), and from the DFG through the project DQUANT (project-id 499347025).
\end{acknowledgments}

\appendix

\section{Quantum many-body scars in the PXP model}\label{app:pxp_detail}

\begin{figure}
        \centering
	\includegraphics[width=1.\columnwidth]{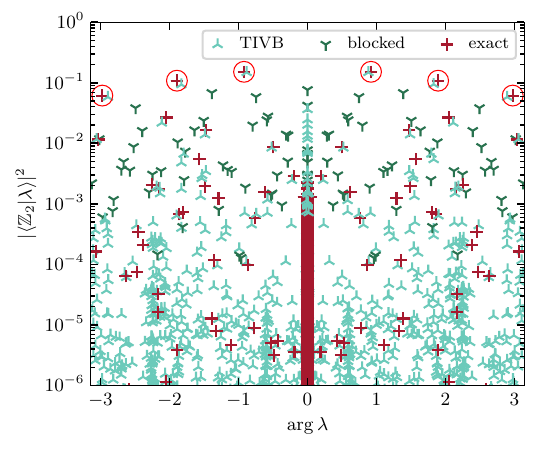}
\caption{The absolute overlap of the eigenstates $\ket{\lambda}$ with the N\'eel state $\ket{\mathbb{Z}_2}$, versus the argument of the complex eigenvalues $\lambda$, for the exact PXP time-evolution operator and compressed circuits with $M=16$ layers at time $t=4$ for system size $L=14$. Note that these are the eigenstates of the entire unitary, not only the first block. Even though the TIVB circuit simulates the time evolution most accurately, the breaking of the constraint leads to multiple additional non-degenerate eigenstates with small overlap. The six exact eigenstates with the largest overlap are encircled in red. The TIVB architecture can reproduce them to high precision.}
\label{fig:4}
\end{figure}

The PXP model is known to host so-called quantum many-body scars~\cite{bernien2017probing,quantum_turner_2018}. In this section we probe the accuracy with which these states are reproduced by the different architectures.
To do so, we show in Fig.~\ref{fig:4} the absolute overlap between the eigenstates $\ket{\lambda}$ of the circuits with eigenvalues $\lambda$ and the $\mathbb{Z}_2$ N\'eel state. The calculations were performed at $L=14$ for $t=4$. 
The overlap with eigenstates of the blocked circuit with $M=16$ are shown in dark blue, and of the TIVB circuit with $M=16$ in cyan. As a reference we show the overlap for the exact time evolution with red crosses. 
The TIVB circuit reproduces the uppermost scarred eigenstates to high accuracy, while the blocked circuit is visibly off. This is in line with their distance, namely $\epsilon\sim\mathcal{O}(10^{-3})$ for the TIVB circuit and $\epsilon\sim\mathcal{O}(10^{-2})$ for the blocked circuit, as can be seen in Fig.~\ref{fig:1}(a). On account of breaking the constraint, there are many spurious states for the TIVB circuit. It is shown in Fig.~\ref{fig:1}(b) that these do not significantly affect the imbalance revivals.

\section{Restricted distances}\label{app:restricted_distances}

\begin{figure}
        \centering
	\includegraphics[width=1.\columnwidth]{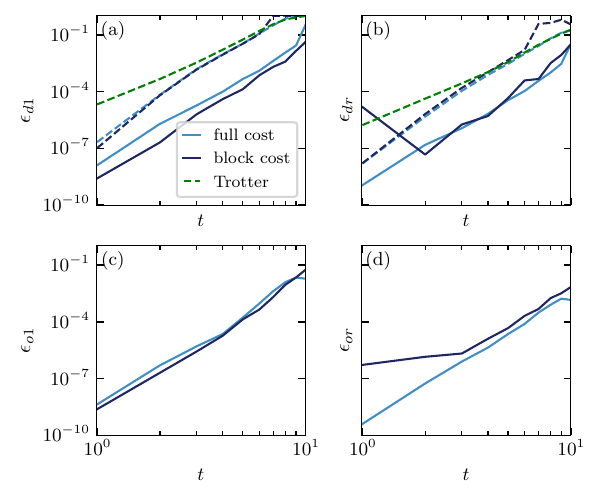}
\caption{The normalized Frobenius distance for various parts of the PXP time-evolution operator with size $L=12$. We consider a TIVB and a blocked circuit with $M=24$, shown as solid and dashed lines, respectively. The second-order Trotter circuit with $M=26$ is shown as a dashed green line. The light blue curves correspond to circuits that were optimized with the total distance $\epsilon$ as cost function, which for the dark blue curves was restricted to the largest diagonal block and its off-diagonal rectangles.
In panel (a) we show the distance of the largest diagonal block $\epsilon_{\textrm{d1}}$, in (b) the distance of the remaining diagonal blocks $\epsilon_{\textrm{d1}}$, in (c) the distance of the off-diagonal rectangles adjacent to the largest diagonal block $\epsilon_{\textrm{o1}}$, and in (d) the distance of the remaining off-diagonal rectangles $\epsilon_{\textrm{or}}$. The block cost function provides a marginal increase in accuracy for the first diagonal block, which is outweighed by the exponential complexity of determining the block. Interestingly, for $t>1$ the restricted cost function achieves equally high accuracy outside of the first block.}
\label{fig:5}
\end{figure}

To better understand how the TIVB circuits can outperform the blocked circuits in compressing the PXP time-evolution operator, we plot in Fig.~\ref{fig:5} the Frobenius distance (\ref{eq:distance}) restricted to various parts of the unitary. We do this for a TIVB and a blocked circuit, both with $M=24$ brickwall layers, shown as the solid and dashed light blue lines. The second-order Trotter circuit with $M=26$ is shown as a dashed green line.
In panel (a) we show the distance of the largest diagonal block $\epsilon_{d1}$, in panel (b) the distance of the remaining diagonal blocks $\epsilon_{dr}$, in panel (c) the distance of the off-diagonal rectangles adjacent to the largest block $\epsilon_{o1}$, and in panel (d) the distance of the remaining distances. To normalize the Frobenius norm we divide it by $2\sqrt{N_{\textrm{e}}}$, with $N_{\textrm{e}}$ being the amount of matrix entries in the average. This reduces to the normalized distance (\ref{eq:distance}) when the average is over one or multiple diagonal blocks.

We see that all distances are roughly equal, indicating that optimizing the total $\epsilon$ leads to systematic improvement of all sectors. The TIVB circuit has almost two orders of magnitude higher accuracy on the diagonal blocks than the blocked circuit. We know from Sec.~\ref{sec:distance_and_imbalance} that this gain carries over to the simulation accuracy of the imbalance revivals, even at times when the largest diagonal block couples to the other diagonal blocks with $\epsilon_{o1}\sim\epsilon_{or}\sim\mathcal{O}(10^{-2})$.

Given these results, it is sensible to ask whether we can increase the accuracy of the largest block by restricting the cost function to the largest block and the adjacent off-diagonal rectangles. The results of this optimization are shown as the dark blue lines in Fig.~\ref{fig:5}. First we consider the TIVB circuit. There is a marginal improvement in the distance of the largest diagonal block $\epsilon_{d1}$ and the distance of its off-diagonal rectangles $\epsilon_{o1}$. Furthermore, the distance of the other diagonal blocks $\epsilon_{dr}$ and the distance of its off-diagonals $\epsilon_{or}$ are automatically reproduced to almost the same extent as when using the full cost function. Only at $t=1$ the accuracy of the distances $\epsilon_{dr}$ and $\epsilon_{or}$ is diminished, which does not affect the distances $\epsilon_{d1}$ and $\epsilon_{o1}$. We have checked that the imbalance revivals are reproduced equally accurate with both cost functions. For the blocked circuit there is no gain at all, with the differences stemming from the optimization procedure. In summary, there seems to be little merit to a restricted cost function, especially because determining the blocks is exponentially complex in $L$ and requires a priori knowledge of the particular structure of $U$.

\begin{figure}
        \centering
	\includegraphics[width=1.\columnwidth]{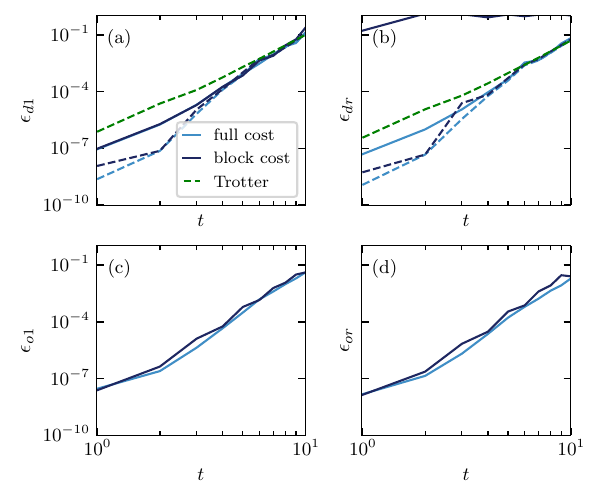}
\caption{The normalized Frobenius distance for various parts of the XXZ time-evolution operator with size $L=12$. We consider a TIVB circuit with $M=24$ and a blocked circuit with $\tilde{M}=8$, shown as solid and dashed lines, respectively. The second-order Trotter circuit with $\tilde{M}=8+1/2$ is shown as a dashed green line. The light blue curves correspond to circuits that were optimized with the total distance $\epsilon$ as cost function, which for the dark blue curves was restricted to the largest diagonal block and its off-diagonal rectangles.
In panel (a) we show the distance of the largest diagonal block $\epsilon_{\textrm{d1}}$, in (b) the distance of the remaining diagonal blocks $\epsilon_{\textrm{d1}}$, in (c) the distance of the off-diagonal rectangles adjacent to the largest diagonal block $\epsilon_{\textrm{o1}}$, and in (d) the distance of the remaining off-diagonal rectangles $\epsilon_{\textrm{or}}$. The block cost function does not improve the accuracy of the first diagonal block, and it no longer automatically yields high accuracy on the smaller diagonal blocks.}
\label{fig:5_xxz}
\end{figure}

In Fig.~\ref{fig:5_xxz} we repeat this analysis for the XXZ model, now for a TIVB circuit with $M=24$ brickwall layers of CNOT gates and a blocked circuit with $\tilde{M}=8$ brickwall layers of $U(1)$-symmetric gates. The second-order Trotter circuit has $\tilde{M}=8+1/2$ brickwall layers of $U(1)$-symmetric gates, with the $+1/2$ indicating an extra half brickwall layer. Clearly, the restricted cost function provides no benefit.

\bibliography{references}

\end{document}